\documentclass[preprintnumbers,prd,onecolumn,showpacs,floatfix,nofootinbib,
superscriptaddress]{revtex4}
 \pdfoutput=1
\usepackage{graphicx}
\usepackage{epsfig}
\usepackage{bm}
\usepackage{amssymb}
\usepackage{float}
\usepackage{amsmath}
\usepackage{dcolumn}
\usepackage{cancel}
\usepackage[colorlinks]{hyperref}
\usepackage[usenames,dvipsnames]{color}
\hypersetup{
     breaklinks=true,
    pdfstartview={FitH},    % fits the width of the page to the window
    colorlinks=true,       % false: boxed links; true: colored links
    linkcolor=blue,          % color of internal links
    citecolor=red,        % color of links to bibliography
    filecolor=magenta,      % color of file links
    urlcolor=blue,           % color of external links
    anchorcolor=green,      % Color for anchor text
    linktocpage=true
}

\def\doi{http://doi.org}

\begin{document}

\title{ Building healthy Lagrangian theories with machine learning}

\author{Christos Valelis } 
\affiliation{Department of Informatics \& Telecommunications, National \& 
Kapodistrian University of Athens, 
Zografou Campus GR 157 73, Athens, Greece}

\author{Fotios K. Anagnostopoulos }
%\email{fotis-anagnostopoulos@hotmail.com}
\affiliation{Department of Physics, National \& Kapodistrian University of Athens, 
Zografou Campus GR 157 73, Athens, Greece}

\author{Spyros Basilakos} 
\affiliation{Academy of Athens, Research Center for Astronomy and
Applied Mathematics, Soranou Efesiou 4, 11527, Athens, Greece}
\affiliation{National Observatory of Athens, Lofos Nymfon, 11852 Athens, 
Greece}

\author{Emmanuel N. Saridakis}
\email{msaridak@phys.uoa.gr}
\affiliation{National Observatory of Athens, Lofos Nymfon, 11852 Athens, 
Greece}
\affiliation{Department of Physics, National Technical University of Athens, Zografou
Campus GR 157 73, Athens, Greece}
\affiliation{Department of Astronomy, School of Physical Sciences, University of Science 
and Technology of China, Hefei 230026, P.R. China}


%%%%%%%%%%%%%%
\begin{abstract}
 The existence or not of pathologies in the context of 
Lagrangian theory is studied with the 
aid of Machine Learning algorithms. Using an example in the framework of 
classical mechanics, we make a {\it proof of concept}, that   the construction of new 
physical theories using machine learning is possible. Specifically, we utilize 
a fully-connected, feed-forward neural network architecture, aiming to 
discriminate between ``healthy'' and ``non-healthy'' Lagrangians, without 
explicitly extracting the relevant equations of motion. The network, after 
training, is used as a fitness function in the concept of a genetic algorithm 
and new healthy Lagrangians are constructed. These new Lagrangians are 
different from the Lagrangians contained in the initial data set.  Hence, 
searching for Lagrangians possessing a number of pre-defined properties is 
significantly
simplified within our approach.
The framework employed in this work can be used to explore more complex physical 
theories, such as  generalizations of General Relativity in gravitational 
physics, or constructions in solid state physics, in which the standard 
procedure can be laborious. 
\end{abstract}

\maketitle

\section{Introduction}

After a century of theoretical research in the construction of physical 
theories, we know that the action, and thus the corresponding  Lagrangian 
density, is the cornerstone quantity, since it gives rise to the equations of 
motion  by employing the least-action principle. During the last century rapid 
progress in the direction of a unified description of physics took place, and 
the combined description of electromagnetic, weak and strong nuclear forces 
within the standard model of particle physics Lagrangian is established. 
However, the inclusion of gravity in this picture is notoriously difficult, 
resulting to a number of different approaches (see \citep{woodard2009far, 
smolin2003far} for a review and  \citep{callender2001physics} for a wider 
perspective of the problem).

One interesting approach is to study higher-order 
corrections to the Einstein-Hilbert action of general relativity, which 
is a sufficient condition to   construct 
  renormalizable and thus potentially quantizable gravitational theories 
\citep{buchbinder2017effective}.  A serious drawback in this approach is that 
including higher derivatives in the Lagrangian is known to cause problems, 
such as the existence of unphysical states, the so-called ghosts. Thus, one is 
indeed interested in Lagrangians that posses terms with higher-derivatives, but 
which  are ghost-free and have equations of motion that are of second order, 
which are generally called ``healthy'' Lagrangians. Searching for healthy 
Lagrangians is usually done by hand, or at most by using a computer algebra 
system, and is a tedious procedure which requires the explicit derivation of 
field equations and the examination  of whether they are of second order.

The ubiquity of massive data volumes, coupled with scalable training techniques, 
novel learning
algorithms and immense computational power, has been an important factor for the 
practical success
of (deep) machine learning (ML). From a predictive accuracy standpoint, deep 
learning algorithms are
nowadays considered the state of the art in numerous applications, while deep 
learning's success has
been the primary reason for a fresh look at the transformative potential of 
artificial inteligence (AI) as
a whole during the past decade.
This success has also made a long-standing dream, that of automated, data-driven 
scientific
discovery, seem within reach. From early approaches in biology and the life 
sciences, \citep{king2011formalization}, to more
recent ones in a wide range of scientific fields \citep{waltz2009automating}, 
AI-assisted scientific discovery is pursued
towards understanding experimental findings, inferring causal relationships from 
observational data,
or even acquiring novel scientific insights and building new explanatory 
theories.
At the same time, however, and despite recent successes and optimism, a 
transparency barrier is
imposing severe limitations on the applicability of the AI potential in 
scientific discovery. It is now
widely acknowledged that the inability to understand, explain and trace back to 
a cause the
predictions/inferences of black-box machine learning models is responsible for 
their lack of
accountability and for end user's lack of trust in such models in 
mission-critical applications and
high-stakes decisions. In fields where understanding, as opposed to merely 
predicting, is the basic
requirement, such as that of scientific discovery and automated explanatory 
theory building, the
opaqueness of state-of-the-art machine learning is a serious handicap.

On the other hand, a consensus is emerging within the machine learning community 
for the existence
of a trade-off between interpretability and predictive performance, especially 
in problems featuring
highly unstructured data sources, as is often the case in natural sciences. In 
such cases, the inner
workings of simpler machine learning models are transparent at the cost of lower 
performance, while
highly complex models are more accurate, at the cost of interpretability. 
Explainable AI, \citep{adadi2018peeking}, attempts
to shed light into the complexities of state-of-the-art machine-generated models 
without
compromising their performance. It is attracting significant attention and 
numerous approaches have
already been proposed, involving e.g. the use of focus mechanisms in deep neural 
networks to
highlight regions of the network involved in certain inferences and potentially 
identify relevant
features, \citep{wagner2019interpretable}, or by identifying such features via 
studying the effects of input perturbations to the
output. Other approaches rely on dissecting the object (e.g. an image) to be 
classified into a set of
prototypical parts and then combining evidence from such prototypes to explain 
the final
classification, \citep{choi2016retain}. The idea behind this paper is to 
automate the search for a healthy Lagrangian possessing higher-derivatives using 
Machine Learning (ML).

In this paper, we are interested in studying  how efficient an artificial neural 
network can be in classifying a given Lagrangian to ``healthy'' or 
``non-healthy'', after training, \emph{without} extracting the equations of 
motion. Further, we construct \emph{new} Lagrangians in an automated way.  The 
manuscript is organized as follows. In Section \ref{Description} we 
provide a concise presentation of the Lagrangian formalism and we also present 
our algorithms for creating Lagrangians. In Section \ref{Algorithms} we describe 
in detail the ML algorithms and architectures we employ, while in Section 
\ref{Discussion} we present and discuss our results. Finally, in Section 
\ref{Conclusions} we draw our conclusions and we point out directions for 
further work.

\section{Description of physical theories}
\label{Description}

A framework that allows systematical study of a physical theory is the 
Lagrangian formalism. The latter comes from classical physics and it was first 
formulated from Joseph-Louis Lagrange in late 18th century. Along with its 
generalizations to include various fields and curved space-time    is the 
cornerstone of modern and classical physics. 
Let us begin our discussion within the context of classical mechanics. Given the 
action \emph{functional}
\begin{equation}
\label{actions}
    \mathcal{S}[\bold{q},t_1,t_2] = 
\int_{\bold{q}(t_1)}^{\bold{q}(t_2)}\mathcal{L}(\bold{q},\dot{\bold{q}},...,
\bold{q}^{(n)})dt,
\end{equation}
where $t_{1}, t_{2}$ are two time instances and $\bold{q}$ is the fundamental 
quantity that describes our physical system and specifically the position over 
time of the particle under study. The corresponding equations of motion arise by 
applying the Hamilton's principle, namely the evolution of the physical system 
is such a way that the action is stationary (for a thorough presentation of 
action principle see \cite{goldstein2002classical}). In other words, the 
trajectory 
that the particle will follow between $\bold{}q(t_1)$ and $\bold{}q(t_2)$ is 
such that the action functional derivative is zero, i.e. $\delta S  = 0$. This 
requirement produces the equation of motion   \cite{goldstein2002classical}
\begin{equation}
\label{Euler_LagrangeEQs}
    \frac{\partial \mathcal{L}}{\partial \bold{q}} -\frac{d}{dt}\left( 
\frac{\partial \mathcal{L}}{\partial \bold{\dot{q}}} \right) + ... + (-1)^{n} 
\frac{d^{n}}{dt^{n}}\left( \frac{\partial \mathcal{L}}{\partial \bold{q}^{(n)} } 
\right) = 0.
\end{equation}
In the latter we omitted the arguments of the Lagrangian for brevity. The 
Lagrangian to obtain the Newtonian mechanics is $\mathcal{L} = T - V(\bold{q})$, 
where $T$ is the kinetic energy of the system, i.e. $T = (1/2) \ m \ 
\bold{\dot{q}}^2$. 
Since the late 18th century where this formalism was first proposed, it has 
been 
generalized to include classical and quantum fields, as well as to handle in a 
unified way time and positions in the context of Einstein's theory of gravity 
\citep{maggiore2005modern, weinberg1972gravitation,Saridakis:2021lqd}. Today the 
procedure of 
constructing a new physical theory is essentially the process of constructing 
new Lagrangians with the desired properties (e.g. specific symmetries).

\subsection{Ostrogradsky's instability}

As a starting point in constructing a Lagrangian, one can consider any function 
of positions and velocities that is smooth enough. In this procedure there 
exists a number of theoretical requirements, such as the verification of 
specific symmetries. However, as Ostrogradsky showed first 
\citep{ostrogradsky1850memoires}, there exists an additional powerful constraint 
that does not allow Lagrangians with higher order time-derivatives. This 
requirement arises from the fact that in such case the corresponding 
Hamiltonian 
is  not bounded, which implies that the total energy of the system is unbounded.

In order to avoid this Ostrogradsky instability the equations of motion of a 
physical system need to have up to second order equation of motion. A sufficient 
condition for that is if the Lagrangian has terms up to first order in 
derivatives, nevertheless this is not necessary since as Horndeski showed 
\citep{Horndeski:1974wa} one may have higher-order terms in a Lagrangian in a 
suitable combination that the resulting higher-order time derivatives in the 
equations of motion cancel out, resulting to second-order ones. Hence, this 
implies that discriminating if a given Lagrangian is healthy or not, requires 
the extraction of the equations of motion and their subsequent elaboration, a 
procedure that in case of complicated Lagrangians is not trivial. That is why it 
becomes of great interest the construction of a machine learning procedure that 
could discriminate if a given Lagrangian is healthy or not without the need to 
perform the variation and the extraction of equations of motion.

\subsection{Describing a Lagrangian}

A Lagrangian could in principle contain scalar quantities that are functions of 
higher dimensional objects. For instance the electromagnetic Lagrangian 
contains 
the tensor $F_{\mu\nu}$, while the Einstein-Hilbert Lagrangian of general 
relativity possesses contractions of the Riemann tensor. In order to give a 
Lagrangian as input to a neural network it is necessary to translate it in a 
form that the network can handle, i.e to describe it as a set of features. 
Formally, a function $\phi: \mathcal{L} \rightarrow \mathbb{R}^n$ is to be found in order 
to provide the feature vectors, where $\mathcal{L}$ is the Lagrangian's space. 

We limit ourselves to classical mechanics, in a sense that the fundamental 
quantity involved in the Lagrangian is the position of a moving body, $x(t)$ 
as a function of time. Position along with its time derivatives constitute the 
set of physically interesting entities. Within our modelling, all physical 
parameters such as mass are set to unity. Moreover, the parameters are measured 
in proper units, in order  to maintain the standard physical interpretation 
of 
potential energy. Some simple representations are provided in the following.

\begin{itemize}
    \item A first parametrization is to use the kinetic term $T = \alpha_0 \ m 
x(t)^{\alpha_{1}}\dot{x}(t)^{\alpha_{2}}\ddot{x}(t)^{\alpha_{3}}$, which 
includes the   standard kinetic term  $T = m 
\dot{x}(t)^2/2 $. Additionally, we consider  a   general potential of the form
    \begin{equation}
    \label{caseA}
        V_{a_1}(x(t), \dot{x}(t), \ddot{x}(t)) = 
\alpha_{4}x(t)^{\alpha_{5}}\dot{x}(t)^{\alpha_{6}}\ddot{x}(t)^{\alpha_{
7}}.
        \end{equation} 
    Hence, the corresponding set of numbers to describe this Lagrangian is 
$\bold{f} = [\alpha_0,...,\alpha_7]$. Using the standard expression for the 
kinetic energy implies that $[\alpha_0,\alpha_1,\alpha_2,\alpha_3] = 
[0.5,0,2,0]$. By using eq. \eqref{Euler_LagrangeEQs} it is easy to observe 
that the essential requirement for a Lagrangian of the form \eqref{caseA} to be 
free of the Ostrogradski instability is to have $\alpha_{4} = 0$ or $\alpha_{4} 
= 1$.
    The aforementioned representation could generalized by considering a sum of 
potentials of the form of \eqref{caseA}:
    \begin{equation}
       \label{caseA22}
         V_{a_2}(x(t), \dot{x}(t), \ddot{x}(t)) = \sum_{i=1}^{n} 
\alpha_{i0}x(t)^{\alpha_{i1}}\dot{x}(t)^{\alpha_{i2}}\ddot{x}(t)^{
\alpha_{i3}},
    \end{equation}
    with $i=1,\cdots,n$,
    and the corresponding feature vector to describe this Lagrangian is  
$\bold{f} = [0.5,0.0,2,0.0,..,\alpha_{n0},\alpha_{n1},\alpha_{n2},\alpha_{n3}]$, 
with dimension $4(n+1)$. 
    
    \item Another parametrization is to further assume that every potential 
that 
describes a certain class of phenomena will be $C^{\infty}$, namely 
 infinite times differentiable, at least within a 
certain subspace. From the latter, the corresponding Taylor series always 
exists, thus an adequate description of a general potential energy is its Taylor 
coefficients around the point  $(x_{0},\dot{x}_{0},\ddot{x}_{0})$. The 
most general Taylor expansion reads
    \begin{equation}
       \label{caseB22}
         V_{b}(x, \dot{x}, \ddot{x}) = \left. \sum_{j = 0}^{\infty} 
\left\{\frac{1}{j!} \left((x - x_{0}) \frac{\partial}{\partial {x}'} 
+(\dot{x} - \dot{x}_{0}) \frac{\partial}{\partial \dot{x}'} + 
(\ddot{x} - \ddot{x}_{0}) \frac{\partial}{\partial \ddot{x}'} 
\right)^{j}V(x', \dot{x}', \ddot{x}') \right\}\right|_{(x' = 
x_0,\dot{x}' = \dot{x}_{0},\ddot{x}' = \ddot{x}_{0)}}.
    \end{equation}
    By keeping terms up to 2nd order and re-arranging we obtain the following
    \begin{equation}
      V_{b_1}(x, \dot{x}, \ddot{x}) = a_{0} + a_{1}x + a_{2} 
\dot{x} + a_{3}\ddot{x} + a_{4}{x}^2+ a_{5}\dot{x}^2 + 
a_{6}\ddot{x}^2  + \alpha_{7} x \dot{x} + \alpha_{8} x \ddot{x} + 
\alpha_{9} \dot{x} \ddot{x} ,
    \end{equation}
    where $a_{i}$ are real numbers and the corresponding feature vector is 
$\bold{f}=[0.5,0.0,2.0,0.0,\alpha_{0},..,\alpha_{9}]$. The terms $\alpha_0$ 
and 
$\alpha_{9}$ do not play any role in a Lagrangian formulation (their 
contribution to the field equations is zero), while the governing parameter is 
$\alpha_{6}$. By including terms up to the 3rd order, we obtain the potential 
$V_{b2}$ as
    \begin{eqnarray}
  &&  
  \!\!\!\!\!\!\!\!\!\!\!\!\!\!
  V_{b_2}(x, \dot{x}, \ddot{x}) = a_{0} + a_{1}x + a_{2} 
\dot{x} + a_{3}\ddot{x} + a_{4}{x}^2+ a_{5}\dot{x}^2 + 
a_{6}\ddot{x}^2  + \alpha_{7} x \dot{x} + \alpha_{8} x \ddot{x} + 
\alpha_{9} \dot{x} \ddot{x}\nonumber\\
&&
\ \ \ \ \ \ \ \ \ \ \ 
 + \alpha_{10}x^3 + 
\alpha_{11}\dot{x}^3 + \alpha_{12}\ddot{x}^3 + \alpha_{13}x \dot{x} 
\ddot{x} + \alpha_{14} x^2 \dot{x} + \alpha_{15} \dot{x}^2 x + 
\alpha_{16}\ddot{x}^2 x + \alpha_{17}\ddot{x}\dot{x}^2 .
    \end{eqnarray}
    Similarly to the previous example, we mention  that the term with 
coefficient $\alpha_{17}$ is a total 
derivative and thus has no effect in the equations of motion. Finally, note 
that in principle, a series expansion could be performed on different 
bases, i.e Legendre, Laguerre and Hermitte polynomials. 

\end{itemize}

In this work, without loss of generality we will use only parametrization 
$V_{a_i}$, since we are interested in studying the  training of a neural network 
to discriminate between ``healthy'' and ``non-healthy'' theories.

 There exist numerous descriptions with regard to certain classes of 
Lagrangians (finding a description valid for an arbitrary Lagrangian 
is 
left for a future project). In Table \ref{lagrangians} some classical 
potentials
are presented along with the proper map to construct the relevant feature 
vectors.

\begin{table}[ht]
\tabcolsep 4.0pt
\vspace{1mm}
\begin{tabular}{cccc} \hline \hline
No & $V$ & Parametrization(s) & Description 
 \vspace{0.05cm}\\ \hline
 1 & 0 & $a_1,a_2,b$ & free particle \\
2 & $\frac{1}{2}k r^2$& $a_1,a_2,b$ & mass connected to an ideal spring of 
spring constant 
$k$\\
3 & $\frac{Gm}{r}$ & $a_1,a_2,b$ &   gravitational potential 
 \\
4 &  
$4\epsilon\left[\left(\frac{\sigma}{r}\right)^{12}-\left(\frac{\sigma}{r}
\right)^6\right]$ & $a_2,b$ &  Lennard-Jones potential 
\citep{10.1088/978-1-6817-4417-9ch4} \\
5 & $\epsilon \left(\frac{\sigma}{r}\right)^{\nu}$ & $a_1,a_2,b$ & soft-sphere 
potential   \citep{10.1088/978-1-6817-4417-9ch4}\\ 
6 & $mg\text{cos}(r)$ & $b$ & non-linear oscillator \\
7 & $D\left[e^{-2\alpha(r-r_0)} - 2e^{-\alpha(r-r_0)} \right]$ & $b$ & Morse 
potential  \citep{10.1088/978-1-6817-4417-9ch4} \\
\hline\hline
\end{tabular}
\caption{ Some classical potentials, along with their 
representation 
ability in order to assess the generality of the parametrizations employed.  
Parametrizations $a_1$,$a_2$,$b$ correspond  to \eqref{caseA},\eqref{caseA22} 
and \eqref{caseB22} respectively.
\label{lagrangians}}
\end{table}

\section{Model and Algorithms}
\label{Algorithms}

In order to classify a Lagrangian as healthy or not, we are using 
fully-connected feed-forward neural networks and supervised learning. Regarding 
the automated production of new Lagrangians, we employ Genetic Algorithms.

\subsection{Neural Network setup}

\begin{figure}[ht!]
    \centering
    \includegraphics[width=0.5\textwidth]{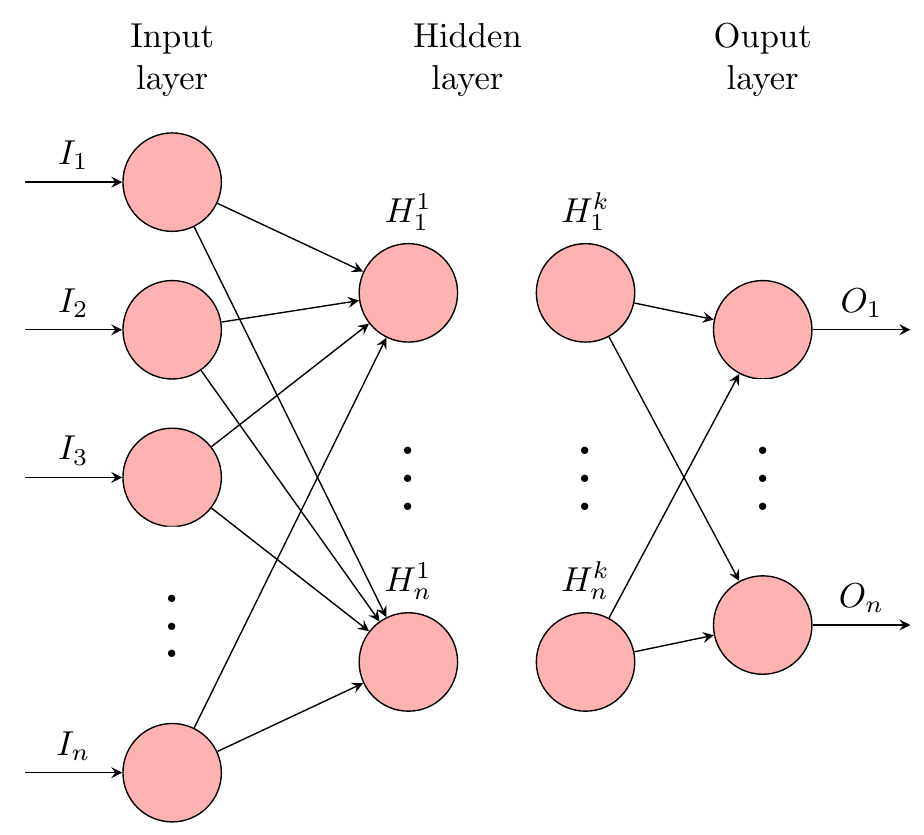}
    \caption{{\it{The structure of a fully - connected neural Network. The 
input and output layers have n neurons, while there are k hidden 
layers, each one consisting of $ n$ neurons.}}}
    \label{fig:NNdepictions}
\end{figure}

A neural network performs mappings from an input space to an output space. In 
our 
case the input space contains the feature vectors of the Lagrangians and the 
output space contains the two categories. The basic structural element of a 
neural network is the \emph{layer}, that is a group of neurons. The neurons of 
each layer connect to those of the next and these connections are called 
\emph{synapses}. The first layer of the network is called input layer, the last 
layer is called output layer and all the layers in between are called hidden 
layers. The architecture mentioned above describes a typical fully connected 
neural network as shown in Fig. \ref{fig:NNdepictions}. The term feed-forward 
implies a network that its synapses do not form a cycle, in contrast with e.g 
the recurrent neural networks  \citep{mandic2001recurrent}. Formal definitions 
of the aforementioned terms regarding feed - forward neural networks could be 
found at
\citep{svozil1997introduction}.

In order to understand how the mapping is realized, we need to break down the 
network to its components. The neurons of the first layer activate as we input 
a 
feature vector (corresponding to a Lagrangian). Activation means that a neuron 
calculate a value and ``feeds'' it to the next layer. As we mentioned above, in 
a fully - connected net, all the neurons of a layer are connected to all 
neurons 
of the previous one through synapses. Each synapse contains a real number, 
called 
weight, that indicates how much the activation of the previous layer specific 
neuron, affects the activation of the neuron in the next layer. Furthermore, 
each neuron contains a real value, called bias, that indicates a threshold, 
over 
which a neuron will meaningfully pass its value over to the next layer (fire). 
The activation of a neuron is described
as
\begin{equation}
\label{activation}
\alpha^{l}_{j} = \sigma \left( 
\sum_{k}w_{jk}^{l}\alpha_{k}^{l-1}+b_{j}^{l}\right),
\end{equation}
where $w_{jk}^{l}$ is the weight of the synapse that connects the $k^{th}$ 
neuron of the $(l-1)^{th}$ layer to the $j^{th}$ neuron in the $l^{th}$ layer. 
Moreover, $b_{j}^{l}$ is the bias of the $j^{th}$ neuron.
The function $\sigma$ is the sigmoid function, $\sigma(x) = 
1/\left(1+\text{exp}(x) \right)$. 

When a neural network is created all its weights and biases values are randomly 
initialized. Therefore, in order for our network to be able to classify 
correctly a given Langragian, its values need to be adjusted.
Specifically these values need  to be adjusted in a way   that when a healthy 
Lagrangian is fed into the network, the network will provide as output a real 
number close to 1, and 
on the other hand in the case of the non-healthy Lagrangian it will provide as 
output a 
number close to 0.

In order to adjust those weights and biases we will employ the back-propagation 
algorithm, described further later in the next subsection.

\subsubsection{Training}

\begin{figure}
    \centering
    \includegraphics[width=\textwidth]{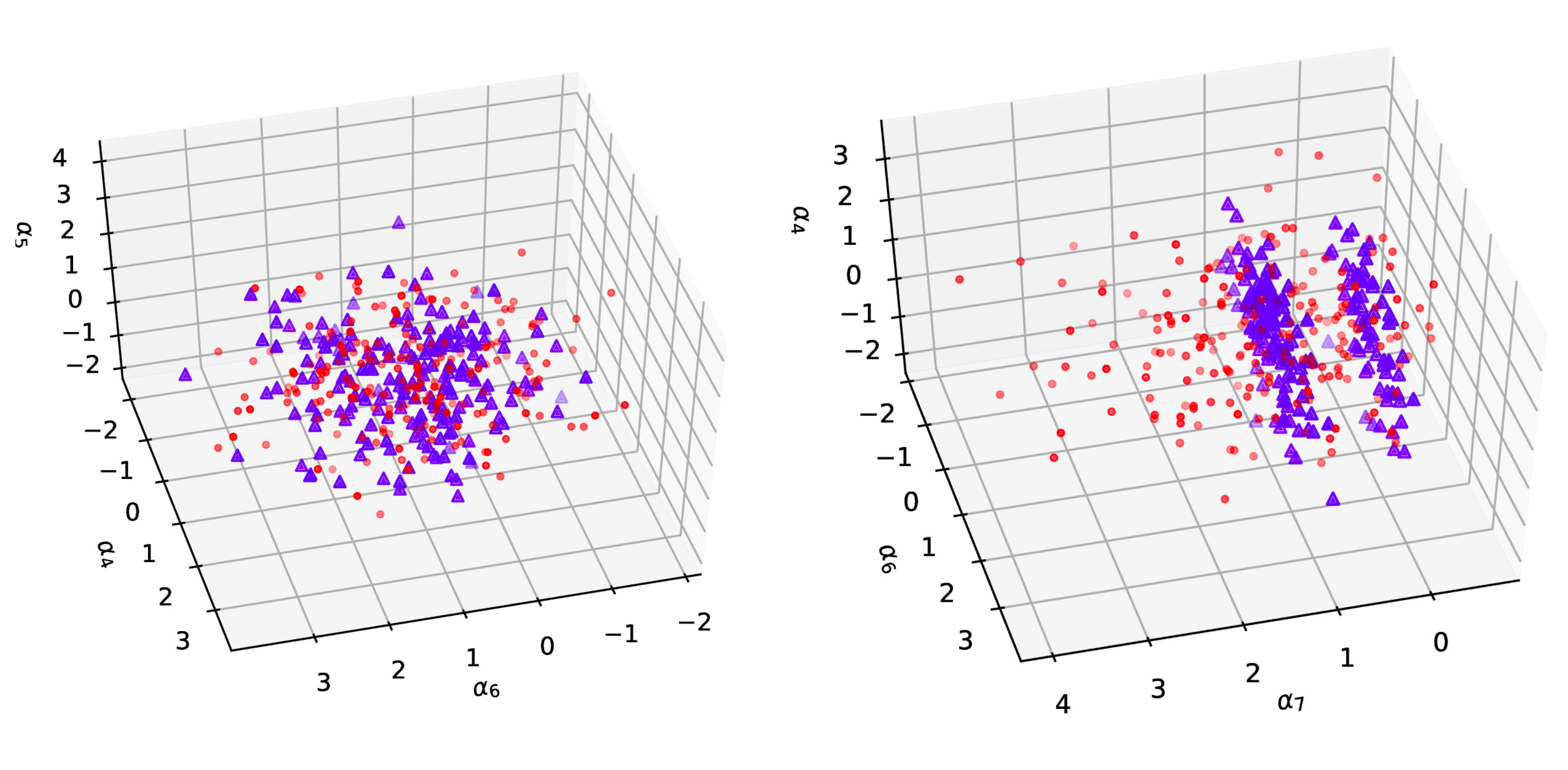}
    \caption{{\it{3-dimensional subspaces of the parameter space for 
Lagrangians of 
the generic form $\mathcal{L} = T - 
\alpha_{4}x^{\alpha_{5}}\dot{x}^{\alpha_{6}}\ddot{x}^{\alpha_{7}}$. The 
circular marker corresponds to ``non-healthy'' Lagrangians while the triangular 
one to "healthy".}}}
    \label{fig:caseA_parameter_space}
\end{figure}

In order to construct the training data set, a Monte Carlo approach is employed.
The i-th iteration of the process consists of constructing a random vector of 
length $n$.
Subsequently, by using a symbolic algebra system, that is \emph{sympy}  
\cite{10.7717/peerj-cs.103},
the Euler-Lagrange equations  \eqref{Euler_LagrangeEQs}  are employed to 
extract 
the equations of motion. Furthermore, the equations are checked for the 
existence of 
time derivatives of order higher than 2. If there are only second derivatives 
in 
the equations of motion, the random vector is labeled as ``healthy'', 
otherwise it 
is labeled as ``non-healthy''. The above procedure is repeated $N$ times. Some 
characteristic subsets of the $a_{i}$ space are presented in Fig. 
\ref{fig:caseA_parameter_space}.  We mention here that  the aforementioned 
criterion of ``healthy'' Lagrangians is definitely not unique. In fact, one 
could choose to label as ``healthy'' any kind of Lagrangian, e.g. Lagrangians 
possessing additional features,   satisfying extra     symmetries, etc. 

The total training data set constitutes of 2000 ``healthy'' and 2000 
``non-healthy'' feature vectors. The training strategy of 
backpropagation  \citep{svozil1997introduction}, 
consists of splitting the total dataset in batches with 400 elements each,
i.e resulting to 10 batches. When all   elements of a batch are fed
into the network, the average cost is computed using the quadratic cost 
function. %, $C = \int (p_{i} - t_{i})$.
The average cost backpropagates through the network fixing its weights and 
biases. 
The same process needs to be done for each batch in order for an epoch of 
training to be completed. 
Before an epoch of training all the training set is scrambled,
resulting in set of different batches. The training object is to minimize the 
cost function.
Different values of the learning rate $\eta$ 
are employed in an effort to avoid local minima, 
while also maintain a steep learning curve. 
All the above steps  describe the iterative process of gradient descend.

\subsubsection{Validation}

After training the network, 10 data sets are produced, 
each of them consisting of 10000 feature vectors, and 
they are given to the network for classification.
In this point another hyper-parameter is introduced, namely the threshold, 
that determines how a number in the continuum set $[0,1]$ 
is projected to the set $[0,1] \cap \mathbb{N}$.
Assessing the ability of the trained network to  correctly classify data that
have not encountered before, is essential in order to avoid over-fitting.
The latter term describes the situation in which the neural network performs 
very well in the training data and poorly in the testing data, 
(for a detailed discussion of over-fitting within a statistical context see 
\citep{geman1992neural}). Towards this purpose, a number of metrics is utilized.

Accuracy measures the relative ability of a
given network to provide \emph{correct} predictions, that is 
\begin{equation}
    \label{Accuracy}
    \text{Accuracy} = \frac{\text{True Positives} + \text{True Negatives}  
}{\text{Total Predictions}}.
\end{equation}
It is straightforward to apply this metric, however there are some caveats.
A characteristic example is that a model that classifies all input data as 
positive,
applied to a dataset consisting of 95 positive and 5 negative samples,
gives 0.95 accuracy score.

Precision is the metric that expresses the proportion of the actual positive 
data  over the data items that were classified by our model as positives:
\begin{equation}
    \label{Precision}
    \text{Precision} = \frac{\text{True Positives}}{\text{True Positives}+ 
\text{False Positives}}.
\end{equation}
Similarly, recall is the metric that expresses what proportion of the actual 
positive data 
was classified correctly by our model:
\begin{equation}
    \label{Recall}
     \text{Recall} = \frac{\text{True Positives}}{\text{True Positives}+ 
\text{False Negatives}}.
\end{equation}
Combining recall and precision by means of the harmonic mean results to the 
$F_1$ 
metric:
\begin{equation}
    \label{F1}
    F_1 = \frac{2\text{Precision} \times \text{Recall}}{\text{Precision} + 
\text{Recall} }.
\end{equation}

Up to this point, the numerical value of a network prediction is not used 
explicitly, since only the classification result is considered.
In order to take into account the numerical value of the network's output, another
metric is defined, namely the Logarithmic Loss. Although in principle one 
could utilize just any function as metric, Log Loss is motivated from information theory, 
and in particular from Kullback-Leibler information \citep{goodfellow2016deep}:
\begin{equation}
    \label{Log Loss}
    \textrm{LogLoss} = -\frac{1}{N} \sum_{i=1}^{N} \sum_{j = 1}^{M} y_{ij} 
\textrm{log}(p_{ij}),
\end{equation}
where        N is the length of the training data set, M the number of different 
classes, $y_{ij}$ the probability
        of the i-th data point to belong to the j-th class, and $p_{ij}$ the 
predicted probability.
Logarithmic Loss does not have upper bound and exists in the range $[0,\infty)$. 
Logarithmic Loss nearer to 0 indicates higher accuracy, 
whereas if Logarithmic Loss is away from 0 then it indicates lower accuracy.

\subsection{Constructing new Lagrangians}

 An intriguing path to explore is the possibility of  constructing new 
Lagrangians   in an automated way, with the aid of the trained network. In this 
work, in order to maintain transparency of the generating procedure, we use 
genetic algorithms (GA). In general there are alternative ways too, e.g.
within the framework of Generative Adversarial Networks 
\cite{goodfellow2014generative}. 
 
 The main idea behind Generative Algorithms is that given a population and a way 
to assess the fitness of each member, if the result of the mating depends on the 
fitness of each parent, after a number of ``generations''  the fitness of the 
population will be maximum. A more formal consideration seems to be in place. 
We assume that a population of feature vectors $P$, with $P \subseteq 
\mathbb{R}^{n}$, exists. 
 The trained network could be considered as a fitness function, $N_{fit}:P 
\rightarrow [0,1]$, and by using it one could associate a value within the 
interval $[0,1]$ to each member of the population. Furthermore, different 
members of 
the population are ``mating'' to create new ones, with the contribution of each 
parent's feature to the corresponding ``child's'' feature to be analogue with 
their 
relative fitness. Lastly, a random ``mutation'' to one or more elements of each 
``child'' occurs according to a pre-defined distribution. These steps 
correspond 
to a ``generation''. After performing the aforementioned steps for a number of 
generations, we have a population whose each member has close to maximum 
fitness. The final population does not have common members with the initial one. 
For more details one could follow the presentation of the subject in 
\cite{goldberg2006genetic}. 

In this work, the       ``mating'' algorithm consists of a 
random selection of pairs within the initial population, and the ``child'' 
possesses
the weighted average of the parent's features. The ``mutation'' step takes 
place 
at a random feature of the child if a dice is smaller that a pre-defined 
probability. The final result obviously is affected by the initial population 
and the mating and mutation operations. The details of the Generative 
Algorithms  are presented in Table \ref{tab:BAOS_data}. 
 \begin{table}[ht]
\begin{tabular}{ccccc}
\hline
 Algorithm & Parameter(s) & Value(s)  \\
 \hline
%  \vspace{4mm}
- & N & 500 \\
- & generations & 1000 \\
Mating & Parents number & 2    \\
Mating & Parent Selection &  $\lambda_i \sim \ $ Uniform$\left(1,\sum_{i=1}^N w_{i}\right)$\\
Mating & Parents Selection & $P_{i} \in \{P:\lambda_{i} -\sum_{j=1}^{i} w_{j} < 0$ \} \\
Mutation & Probability & $p_{mut} \sim \ $Uniform(0,1) \\
Mutation & Threshold & $p_{mut}<0.01$ \\
Mutation & Feature Selection & $i \sim $  Uniform$\left(1,4\right)$ \\
\hline
\end{tabular}
\caption{Parameters used in the Generative Algorithms approach, N is the 
dimension of the population. The weights  $w_i$  are defined as $w_i = 10^5 
N_{fit}(P_i)$, where $N_{fit}$ is the trained neural network and $P_i$ is a 
member of the population.}
\label{tab:BAOS_data}
\end{table}

\begin{figure}[!t]
    \centering
    \includegraphics[width=0.5\textwidth]{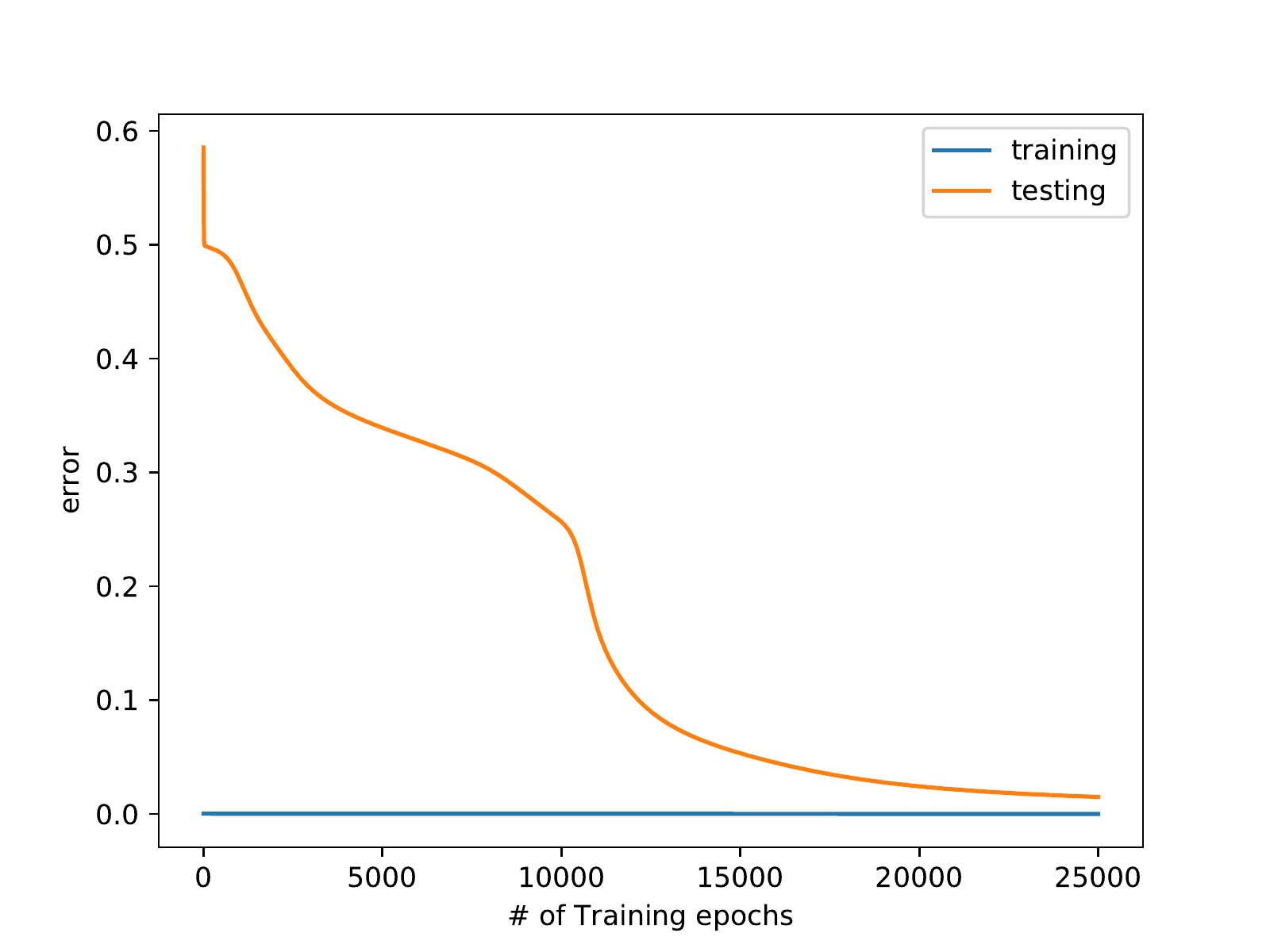}
    \caption{{\it{The training and testing errors as a function of training 
epochs, for 8-6-6-1 network with $\eta = 0.1$.}}}
    \label{fig:ErrorsVsEpochs_1}
\end{figure}

\begin{figure}[!t]
    \centering
    \includegraphics[width=0.5\textwidth]{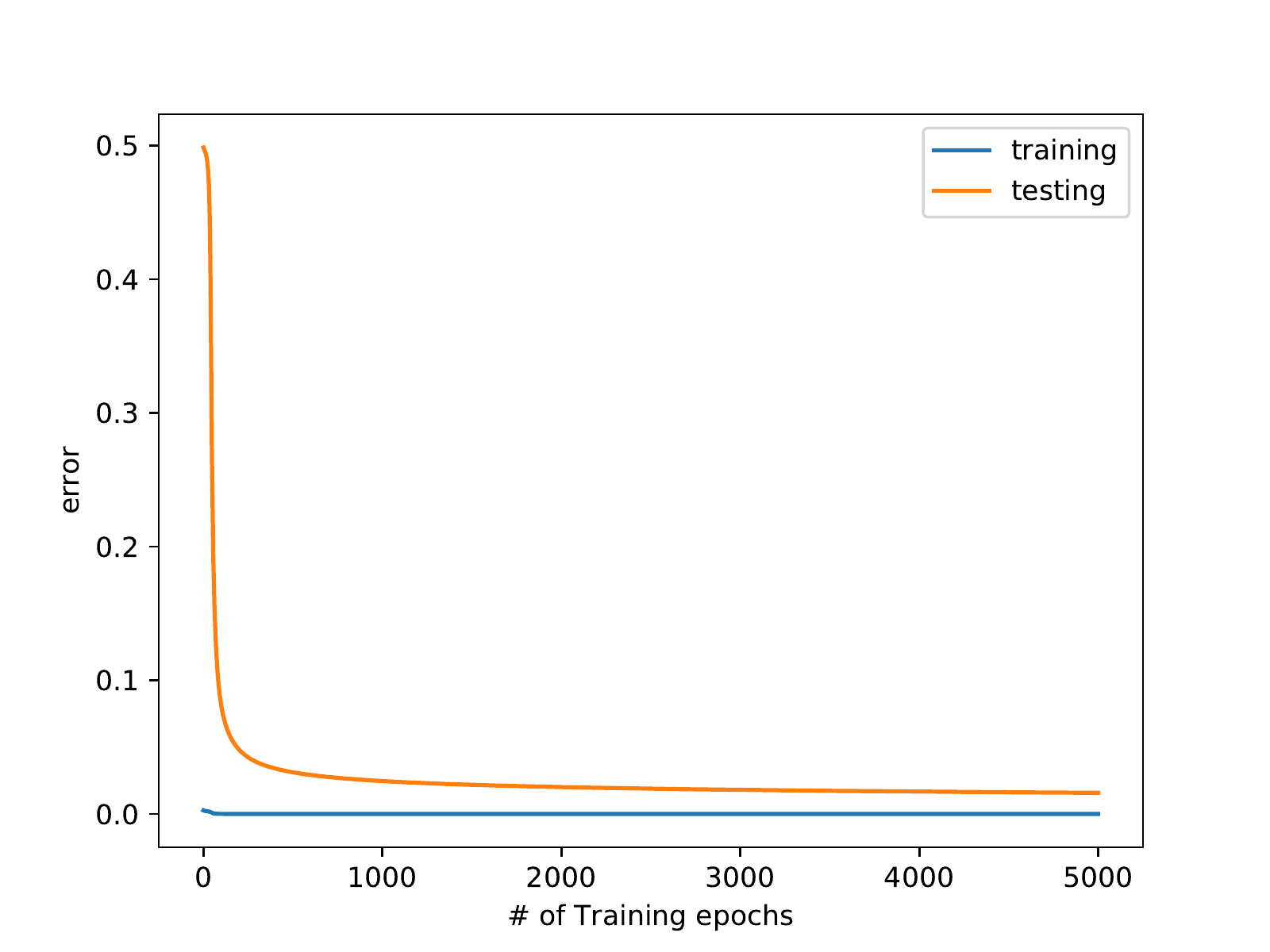}
    \caption{{\it{The training and testing errors as a function of training 
epochs, for 18-16-16-1 network with $\eta = 1$.}}}
    \label{fig:ErrorsVsEpochs_2}
\end{figure}

\section{Results and Discussion}
\label{Discussion}

A fully-connected neural network architecture is implemented and trained to 
discriminate 
between ``healthy'' and ``non-healthy'' Lagrangians for a classical theory
with one degree of freedom. A number of different architectures, regarding the 
depth 
and the width of the network, were tested and the most promising was selected,
namely a structure with two hidden layers of 16 neurons each.
The optimal number of epochs to be used for training the network, in order
to minimize training error on the one hand while   avoiding over-fitting on the 
other,
is found by comparing training and validation errors.
Over-fitting occurs when we observe an increase in the testing error while the 
training error is consistently decreasing (see p.~202 of 
\cite{rao2018computational}). 
However, in our case the testing error seems to be decreasing over
the epochs, resulting in a better ``behaviour'' of the network, as can be seen 
in 
Figs. \ref{fig:ErrorsVsEpochs_1}, \ref{fig:ErrorsVsEpochs_2}. Moreover, we also apply the metrics defined in
Section \ref{Algorithms} at different stages of the training procedure, as can 
be  seen in Tables \ref{tab:Results1} and \ref{tab:Results2}.

\begin{table}[ht]
\tabcolsep 4.0pt
\vspace{1mm}
\begin{tabular}{cccccc} \hline \hline
Epoch & Accuracy & Precision(s) & Recall &$F_1$& Log. Loss
 \vspace{0.05cm}\\ \hline
 500 &  0.549 &  0.081 & 0.580 &  0.564 & 0.682 \\%%ok
5000 & 0.613 &  0.143 & 1.0 & 0.760 & 0.549  \\%%ok
10000 &   0.685 & 0.164 &  0.940 &  0.793 & 0.374 \\ %ok
15000 & 0.965 & 0.653 & 0.999 & 0.982 & 0.067\\ %ok
20000 & 1.000 & 1.000 & 1.000 & 1.000 & 0.026\\%ok
25000 & 1.000 & 1.000 & 1.000 & 1.000 & 0.016\\%ok
\hline\hline
\end{tabular}
\caption{ Values of different metrics during epochs of training for the 
fully-connected, feed-forward neural network, with architecture 8-6-6-1 and 
$\eta 
= 0.1$. Note 
that early epochs are affected by the initial random selection of 
weights.
\label{tab:Results1}}
\end{table}

\begin{table}[ht]
\tabcolsep 4.0pt
\vspace{1mm}
\begin{tabular}{cccccc} \hline \hline
Epoch & Accuracy & Precision(s) & Recall &$F_1$& Log. Loss
 \vspace{0.05cm}\\ \hline
 200 &  0.965 & 0.975 & 0.973 &  0.969 & 0.119 \\%ok
400 & 0.970 &   0.979 &  0.976 & 0.973 & 0.097  \\%ok
600 &   0.972 &  0.981 &   0.976 &  0.974 & 0.091 \\ %ok
800 & 0.972 &  0.983 & 0.976 &  0.974 & 0.088\\%ok 
1000 & 0.973 &  0.983 &  0.976 & 0.974 &  0.086\\%ok 
3000 & 0.974 & 0.987 &  0.974 & 0.974 & 0.080\\%ok 
5000 & 0.975 & 0.988 & 0.975 &  0.975 &  0.076\\%ok 
\hline\hline
\end{tabular}
\caption{ Values of different metrics during epochs of training for the 
fully-connected, feed-forward neural network, with architecture 18-16-16-1 and 
$\eta = 1$. Note 
that early epochs are affected by the initial random selection of 
weights.
\label{tab:Results2}}
\end{table}

\begin{table}[ht]
\begin{tabular}{ccc}
\hline
 \# & $\mathcal{L}- 1/2\dot{x}^2$  \\
 \hline
%  \vspace{4mm}
1 &   $0.15\ x^{-0.43}  \dot{x}^{-3.58}$  \\
 \vspace{0.3mm}
2 &  $ 0.51\ x^{-2.26} \dot{x}^{-0.32}$ \\
\vspace{0.3mm}
3 & $-0.23 \ x^{ -0.09}  \dot{x}^{1.02}\ddot{x} $ \\  
\vspace{0.3mm}
4 & $-0.01 \  x^{ 0.04}  \dot{x}^{1.72}\ddot{x}$ \\
\vspace{0.3mm}
5 &  $  4.122\  x^{ -0.08}  \dot{x}^{ -0.2}$\\
\vspace{0.3mm}
6 & $2.65\  x^{ -0.17} \dot{x}^{-0.35}$ \\
\vspace{0.3mm}
7 & $ 2.511 \ x^{-0.49} \dot{x}^{-0.5}$\\
\vspace{0.3mm}
8 & $ 3.775 \ x^{-0.2} \dot{x}^{ -0.45} $\\
\vspace{0.3mm}
9 & $0.04 \ x^{-0.23}   \dot{x}^{1.38} \ddot{x} $\\
10 & $ 2.1975 \  x^{-0.4}   \dot{x}^{ -0.11}$ \\ 
%  \vspace{4mm}
\hline
\end{tabular}
\caption{A number of ``new'' Lagrangians that correspond to healthy theories, 
constructed  using GA. The procedure was repeated for 10 times and the most 
fitted 
Lagrangian for each iteration was selected.}
\label{tab:res-new_lagrangians}
\end{table}
By employing GA  we  finally construct new healthy Lagrangians. For  indicative 
purposes, in Table \ref{tab:res-new_lagrangians} we present   a list with the 
most prominent Lagrangians, each selected from the final population, after 200 
generations, from 10 different initial populations. These Lagrangians were 
constructed by the machine learning procedure without the extraction of field 
equations. As one can see amongst others, the GA was able to find the correct 
feature, that a ``healthy'' Lagrangian needs to have $\ddot{x}$ to the exponent 
0,1, although no such rule was initially imposed (concerning the other 
  exponents we   could have additionally required them to be integers, however 
we did not do it since it is irrelevant for our analysis). 

One might argue that 
the gain is not significant, since a human can relatively easily write down 
Lagrangians similar to those of Table \ref{tab:res-new_lagrangians}, that will 
not lead to more than second-order terms in the field equations. Indeed, this 
is 
true for this simple subclass of one-dimensional Lagrangians, but it becomes 
laborious for higher dimensionality and more complicated Lagrangians. For 
instance in the case of gravitational theories, such as General Relativity and 
its modifications, where one uses four dimensions and Lagrangians made from 
contractions of the curvature (Riemann) tensor, it is impossible to deduce 
a priori if a Lagrangian leads to equations of motion with  more than 
second-order terms (and thus whether it is physically accepted or not), unless 
these equations are explicitly extracted. Moreover, 
even after the explicit extraction of the equations of motion, a thorough 
elaboration is needed since terms  with  more than 
second-order derivatives could be mutually eliminated through suitable 
integration by parts (see e.g. 
\cite{Horndeski:1974wa,Saridakis:2016ahq,Erices:2019mkd}). In summary, even 
with the help of usual software  the above standard procedure of generating new 
healthy Lagrangians can be laborious and require many months. Hence, we deduce 
that building machine learning tools that can construct new 
healthy Lagrangians without the need to perform all the steps of the standard 
procedure, namely explicitly extract the field equations and thoroughly examine 
and elaborate them, would be extremely helpful for the 
community of gravity, analytical and quantum mechanics,
solid-state, theoretical physics, biology etc.

\section{Conclusions}
\label{Conclusions}

A neural network architecture was implemented, trained and tested, in order to 
decide 
if a given Lagrangian will lead to equations of motion with higher order 
derivatives, \emph{without} explicitly performing the analytical calculations.
Validation of the trained network using the explicit standard procedure
proved that for the simple model under study, the efficient
discrimination of healthy and non-healthy Lagrangians is established.  
Furthermore, ``new'' healthy Lagrangians were constructed in a fully automated 
way. Their properties are related with the initial Monte Carlo generated 
population of Lagrangians and the imposed requirements. Thus, in general one 
can automatically construct new Lagrangians, corresponding to new physical 
theories, possessing arbitrary properties.
By suitably labeling the training data-sets, one can employ different criteria,
resulting to a trained network that could decide for more complex cases.

There are various applications for this kind of approach, ranging 
from gravitational theories to solid state physics.
In all cases, the search for a Lagrangian possessing 
some pre-defined properties (i.e symmetries)
could be substantially simplified in the context of our approach. 
Additionally,  it is interesting  to mention that such an approach might be 
used to 
discriminate models derived from such Lagrangians. 
For instance in \cite{Escamilla-Rivera:2019hqt}
it was proposed a  deep learning tool in order to study the evolution of dark 
energy models, combining two architectures: the Recurrent Neural Networks  
and the Bayesian Neural Networks, since the former is capable of  classifying 
objects while the latter emerges as a solution for problems like over-fitting.  
 Such applications could be useful to confront  with Lagrange-multiplier based 
models (see e.g. \cite{Capozziello:2010uv,Cai:2010zma}) or other gravitational 
models arising from Lagrangian modifications.  Definitely, 
in more complex applications,   the exact form of the map between the Lagrangian 
space and the $n$-dimensional
real space needs to be defined.
Exploring the different ways to ``translate'' a Lagrangian to multi-dimensional 
real space, in an effort for a general description, is deemed very fruitful, 
and 
the present  analysis can stand as its base.

\end{document}